\documentclass[preprint,showpacs]{revtex4}
\usepackage{graphics,graphicx,epsfig}

\begin{document}
\title{Finite-size effects in roughness distribution scaling}
\author{T. J. Oliveira${}^{a)}$ and F. D. A. Aar\~ao Reis${}^{b)}$
\footnote{a) Email address: tiagojo@if.uff.br\\
b) Email address: reis@if.uff.br}}
\affiliation{Instituto de F\'\i sica, Universidade Federal
Fluminense, Avenida Litor\^anea s/n, 24210-340 Niter\'oi RJ, Brazil}

\date{\today}

\begin{abstract}

We study numerically finite-size corrections in scaling relations for roughness
distributions of various interface growth models. The most common relation,
which considers the average roughness
$\langle w_2\rangle$ as scaling factor, is not obeyed in the steady states of a
group of ballistic-like models in $2+1$ dimensions, even when very large system
sizes are
considered. On the other hand, good collapse of the same data is obtained with
a scaling relation that involves the root mean square fluctuation of the
roughness,
which can be explained by finite-size effects on second moments of the scaling
functions. We also obtain data collapse with an alternative scaling relation
that accounts for the effect of the intrinsic width, which is a constant
correction term previously proposed for the scaling of $\langle w_2\rangle$.
This illustrates how finite-size corrections can be obtained from roughness
distributions scaling. However, we discard the usual interpretation that the
intrinsic width is a consequence of high surface steps by analyzing data of
restricted solid-on-solid models with various maximal height differences between
neighboring columns. We also observe that large
finite-size corrections in the roughness distributions are usually accompanied
by huge corrections in height distributions and average local slopes, as well
as in estimates of scaling exponents. The molecular-beam epitaxy model of Das
Sarma and Tamborenea in $1+1$ dimensions is a case example in which none of the
proposed scaling relations works properly, while the other measured quantities
do not converge to the expected asymptotic values. Thus, although roughness
distributions are clearly better than other quantities to determine the
universality class of a growing system, it is not the final solution for this
task.

\end{abstract}

\maketitle

\section{Introduction}
\label{intro}

In order to understand the basic mechanisms of a growth process, it is a common
practice to associate it to a certain universality class \cite{barabasi,krug}.
This is usually done with the calculation of scaling exponents of surface
roughness and assuming that Family-Vicsek scaling is valid \cite{fv} (with a
suitable generalization when anomalous scaling holds). However, huge
corrections to scaling make this calculation difficult in finite-size systems,
both for simulation work and for the analysis of images of real surfaces.
These problems and the interest in finding more universal properties of growth
processes motivated recent studies of other quantities, such as the
distributions of surface roughness \cite{foltin,racz,antal,marinari1}. Most
efforts were devoted to calculation of those distributions in steady states
(roughness saturation), but the comparison of distributions obtained under
conditions that parallel experimental work (the so-called window boundary
conditions) is also a promising tool
\cite{moulinet,rosso,localKPZ,santachiara}.

It is expected that roughness distributions scale as
\begin{equation}
P\left( w_2\right) = {1\over
\langle w_2\rangle} f\left( {w_2\over{\langle w_2\rangle}}\right) ,
\label{deff}
\end{equation}
where $P\left( w_2\right)$ is the probability density of the roughness
$w_2\equiv \overline{h^2} -{\overline{h}}^2$ of a
given configuration to lie in the range $\left[ w_2, w_2+dw_2\right]$, $f$ is
a scaling function, the overbars denote spatial averages and the angular
brackets denote configurational averages. Eq. (\ref{deff}) is usually obtained
in analytical work, such as in the calculation of distributions for
$1/f^\alpha$-noise interfaces \cite{antal}. An alternative scaling relation is
\begin{equation}
P\left( w_2\right) =
{1\over \sigma} g\left( {{w_2-\left< w_2\right>}\over\sigma}\right) ,
\label{defg}
\end{equation}
where 
\begin{equation}
\sigma\equiv {\left( \langle {w_2}^2\rangle - {\langle w_2\rangle}^2
\right)}^{1/2}
\label{defsigma}
\end{equation}
is the root mean square (rms) deviation of $w_2$.

It is reasonable to claim that comparison of full roughness distributions is
better than the calculation of one or two exponents to determine the class of a
growth process. However, these distributions may also be deviated from the
expected scaling relations by finite-size corrections.  For instance,
finite-size data of some models in the Kardar-Parisi-Zhang (KPZ) class 
\cite{kpz} show small deviations from relation (\ref{deff}), while good data
collapse of numerical data for different lattice sizes is obtained with
relation (\ref{defg}) \cite{distrib}. The advantage of a scaling
relation equivalent to (\ref{defg}) was also illustrated in the study of
maximal height distributions (MHD) of $1/f^\alpha$ signals \cite{gyorgyi}.
Moreover, the leading corrections to scaling in MHD of some one-dimensional
solid-on-solid models were analytically predicted and numerically confirmed in
Ref. \protect\cite{schehr}, which clearly shows that such corrections are not
related to low accuracy of simulation data. The presence of huge scaling
corrections may be much worse in distributions of other quantities, such as
height distributions \cite{chin,shim,marinari,kpz2d}. 

For the above reasons, investigation on the failure of finite-size scaling of
roughness distributions is essential to ensure their reliability for comparison
with real or model systems. Moreover, understanding the reasons for the failure
of an scaling relation and the advantages of other ones may also be helpful to
find the form of the main finite-size corrections and to propose alternative
methods of analysis. One particularly important case is KPZ growth in
$2+1$ dimensions, since this class contains many models with large corrections
to scaling of the average roughness, such as ballistic deposition (BD) and Eden
growth, as well as a large number of applications (see e. g. Refs.
\protect\cite{krim,miettinen,tsamouras}).

The aim of this work is to study finite-size corrections in roughness
distribution scaling by analyzing numerical simulation data of various
interface growth models and to discuss the possible relations with other
quantities that characterize those interfaces. The numerical approach is
essential to study the three-dimensional systems considered here, as well as
some two-dimensional systems (the usual systems where analytic work is possible
are Gaussian interfaces).

First, we will study a group of
ballistic-like models where the deviations from relation (\ref{deff}) are much
larger than those previously found in other models \cite{distrib}. This means
that no evidence on their universality class can be obtained by assuming that
scaling relation. On the other hand, good collapse of large systems data is
obtained with Eq. (\ref{defg}), which shows that they are actually in the KPZ
class. Moreover, we successfully propose an alternative scaling relation
accounting for the effect of the intrinsic width, which is a constant correction
term previously included in the scaling of the average roughness
\cite{stauffer,kertesz}. This shows how the main finite-size corrections
may be extracted from roughness distribution scaling. However,
we discard the usual interpretation that the intrinsic width is a consequence
of high surface steps by analyzing data of restricted
solid-on-solid (RSOS) models, where this correction is not present. Instead, we
observe that the deviations from finite-size scaling in the roughness
distributions are usually accompanied by
significant finite-size dependence of dimensionless amplitude ratios of moments
of height distributions, as well as of scaling exponents and average local
slopes. This conclusion is not restricted to KPZ systems, as shown in the
analysis of the molecular-beam epitaxy model of Das Sarma and Tamborenea (DT)
\cite{dt} in $1+1$ dimensions, which is well-known for the slow convergence of
numerically estimated exponents \cite{tamborenea,lanczycki}. Indeed, for
this model none of the proposed scaling relations is able to produce good
collapse of small systems data.

Although we are not able to determine all the reasons for the failure of scaling
relations in a given universality class, we will show how deviations in
different quantities are connected and will suggest alternative methods of
analysis. This may be helpful not only in future work on complex models and
real systems, but also to search for models with minimal finite-size effects,
e. g. following the ideas of Ref. \protect\cite{ghaisas}. Anyway, an important
conclusion is that roughness distributions are actually superior to other
quantities for a reliable search of the universality class of a growing system,
although it is not the final solution for all models.

The rest of this work is organized as follows. In Sec. II we discuss the above
scaling relations and propose an alternative one for cases where intrinsic width
is expected. In Sec. III we present the discrete models analyzed in the
subsequent Sections and the equations defining their universality classes. In
Secs. IV, V and VI, we discuss the finite-size effects in the roughness
distributions, the role of local slopes and the finite-size effects in the
height distributions, respectively. In Sec. VII we summarize our results and
present our conclusions.

\section{Basic properties of scaling functions}
\label{scalingrelations}

If the scaling relation (\ref{deff}) holds, the function $f(x)\equiv \langle
w_2\rangle P(w_2)$ is a function of the variable $x\equiv
w_2/\langle w_2\rangle$. The moment of order $n$ of this function is defined as
$M_f^{\left( n\right)} \equiv \int{x^nf\left( x\right) dx}$, so that the first
moments are given by 
\begin{equation}
M_f^{\left( 0\right)} = 1 \qquad ,\qquad M_f^{\left( 1\right)} = 1 \qquad ,
\qquad M_f^{\left( 2\right)} = \frac{\sigma^2}{{\langle w_2\rangle}^2} +1 .
\label{momentsf}
\end{equation}
This means that any finite-size dependence of the ratio $r\equiv \langle
w_2\rangle/\sigma$ (between the average square roughness and its rms
fluctuation) will lead to a finite-size dependence of the second moment.
However, in the scaling relation (\ref{defg}),
$g(y)\equiv \sigma P(w_2)$ is a function of the variable
$y\equiv \left(w_2-\langle w_2\rangle\right) /\sigma$, so that the first moments
of $g$ are
\begin{equation}
M_g^{\left( 0\right)} = 1 \qquad ,\qquad M_g^{\left( 1\right)} = 0 \qquad ,
\qquad M_g^{\left( 2\right)} = 1 \qquad ,\qquad M_g^{\left( 3\right)} =
\frac{\langle {w_2}^3\rangle}{\sigma^3} -3r-r^3 .
\label{momentsg}
\end{equation}
This means that the finite-size dependence of $r$ do not lead to deviations in
the second moment of this scaling function, leaving possible corrections to the
third moment.

In works based on data collapse methods, the fluctuations in
the first few moments of a scaling function are the ones most easily related to
visual deviations of the plots of data from different system sizes.
Consequently, scaling relation (\ref{defg}) is expected to be better than
(\ref{deff}) by avoiding that finite-size effects are reflected in its second
moment. This was already observed for some KPZ models in $2+1$ dimensions
in relatively small system sizes \cite{distrib}. Recent work on maximal height
distributions of $1/f^\alpha$ signals also illustrated this feature and
stressed the fact that finite-size scaling of the higher cumulants was related
to the deviations \cite{gyorgyi}.

The knowledge of the particular properties of a growth model is certainly
important to guide the proposal of suitable scaling relations. For instance,
this was the case of a recent work on elastic lines in random environments, in
which the scaling relation (\ref{deff}) was generalized to incorporate time and
temperature effects \cite{bustingorry}. Here, we will consider a series of
ballistic-like models whose deposits contain holes and overhangs, among other
steep features, thus our approach is based on properties of related models.

Simulation work on models which produce deposits with holes and overhangs, such
as the Eden model, suggested that the leading correction to the scaling of the
average squared roughness is a constant term
\cite{stauffer,kertesz,tammaro,chavez,moro}. The time evolution of the
average squared roughness, starting from a flat interface, obeys
\begin{equation}
\xi_2 \left( L,t\right) = L^{2\alpha} F\left( tL^{-z}\right) + W_2 ,
\label{fv}
\end{equation}
where $W_2$ is a constant  called the intrinsic width, $\alpha$ is the
roughness exponents and $z$ is the dynamic exponent.
The usual interpretation of  relation (\ref{fv}) is that the first term at the
right-hand side (the original Family-Vicsek relation \cite{fv}) is associated
with fluctuations of all wavelengths, while the intrinsic width is thought to
be related to short wavelength features, represented by overhangs, holes and
high steps at the film surface - see e. g. Refs. \protect\cite{kertesz,
tammaro}.

The present work is only concerned with steady state properties, where $F(x)\to
const$ in Eq. (\ref{fv}) and the saturation roughness [$\langle w_2\rangle =
\xi_2 \left( L, t\to\infty\right)$] scales as
\begin{equation}
\langle w_2\rangle = A L^{2\alpha} + W_2 , 
\label{fvsat}
\end{equation}
with $A$ constant. Large values of $W_2$ are responsible for remarkable
deviations of finite-size estimates of exponent $\alpha$ from its asymptotic
value. However, numerical work on ballistic-like models show that
finite-size corrections to the scaling of $\sigma$ are smaller than the
corrections in $\langle w_2\rangle$ \cite{sigma}.

These results lead us to propose that, for the ballistic-like
models, the roughness of a given interface configuration is a sum of a
fluctuating value, which is
distributed according to the universality class of the model, and a constant
contribution, which is the intrinsic width. Following this reasoning, we
change relation (\ref{deff}) by subtracting a constant from $w_2$
and $\langle w_2\rangle$:
\begin{equation}
P\left( w_2\right) = {1\over {\langle w_2\rangle-C}}
h\left( {{w_2-C}\over{\langle w_2\rangle -C}}\right) .
\label{defh}
\end{equation}
Here $C$ must be viewed as a fitting constant and $h$ is a scaling function.
The first moments of the function $h(z)\equiv \left( \langle w_2\rangle-C\right)
P(w_2)$,  where $z\equiv {{\left( w_2-C\right)}/{\left(\langle w_2\rangle
-C\right)}}$, are
\begin{equation}
M_h^{\left( 0\right)} = 1 \qquad ,\qquad M_h^{\left( 1\right)} = 1 \qquad 
\qquad M_h^{\left( 2\right)} = \frac{\sigma^2}{{\left( \langle w_2\rangle -C
\right)}^2}+1 .
\label{momentsh}
\end{equation}

If our assumptions are correct and $C$ is chosen to match the intrinsic
width $W_2$, then the finite-size dependence of the second moment
$M_h^{\left( 2\right)}$ will be cancelled, or at least it will be greatly
reduced. Consequently, the quality of a data collapse plot will
be similar to that based on Eq. (\ref{defg}). Certainly this
proposal has to be validated by numerical (or eventually analitic) work, and it
is also important to notice that it is independent on the particular
interpretation given to the intrinsic width. 

\section{Discrete growth models and universality classes}
\label{models}

The discrete models analyzed in this paper are the original BD model, the
RSOS model, a conserved RSOS model, the grain deposition models and
the DT model. In all cases, deposition begins with a flat substrate.  

In BD, each incident particle is released from a randomly chosen position above
the deposit, follows a trajectory perpendicular to the substrate and sticks
upon first contact with a nearest neighbor occupied site~\cite{vold,fv}.

In the RSOS model \cite{kk,alanissila}, the incident particle can stick at the
top of a column only if the differences of heights of all pairs of neighboring
columns do not exceed ${\Delta H}_{max}$ after aggregation. Otherwise, the
aggregation attempt is rejected. In this work, results for ${\Delta H}_{max}=1$,
${\Delta H}_{max}=10$ and ${\Delta H}_{max}=20$ will be presented.

The original conserved RSOS model (CRSOS) was proposed in Ref.
\protect\cite{crsos1}, but here we will consider an extension of that model
that has the same symmetries and, consequently, belong to the same universality
class. In this generalized CRSOS model \cite{crsos2}, hereafter simply called
CRSOS, if the column of incidence does not obey the above condition
for aggregation, with ${\Delta H}_{max}=1$, then the incident particle
executes random walks among neighboring columns until finding a position where
it is satisfied and aggregation can occur.

In the $DT$ model \cite{dt}, the arriving particle sticks at the top of the
column of incidence if it has one or two lateral
neighbors at that position. Otherwise, the neighboring columns are consulted and
if one of them satisfies that condition, then the incident particle aggregates
at that point. However, if no neighboring column satisfies the condition, then
the particle sticks at the top of the column of incidence, and if both
neighboring columns satisfy the condition, then one of them is randomly chosen.

The grain deposition models studied here were
introduced in Ref. \protect\cite{graos} for the study of a crossover in local
roughness scaling similar to experimental systems.
They are defined in a simple cubic lattice where the length unit is the
lattice parameter. The grains have cubic shapes and lateral size $l$. They
sequentially incide
perperdicularly to an initially flat substrate, with two of their faces parallel
to the substrate. The incident grain permanently
aggregates to the deposit when its bottom touches a previously aggregated grain
(thus, there is no lateral aggregation). The process is shown in Fig. 1. Here we
will simulate the model with grain sizes ranging from $l=2$ to $l=16$. Despite
the different aggregation rules, the grain deposition models and the BD model
form the group to which hereafter we will refer as ballistic-like models.

The symmetries of the ballistic-like models and of the RSOS models indicate that
they are in the KPZ universality class (see e. g. the discussion in Ref.
\protect\cite{hagston}, in which the continuous description of the growth
process is provided by the KPZ equation
\begin{equation}
{{\partial h}\over{\partial t}} = \nu_2{\nabla}^2 h + \lambda_2
{\left( \nabla h\right) }^2 + \eta (\vec{x},t) .
\label{kpz}
\end{equation}
Here, $\nu_2$ and $\lambda_2$ are constants, $\eta$ is a Gaussian
noise with zero mean and variance $\langle
\eta\left(\vec{x},t\right) \eta\left(\vec{x'},t'\right) \rangle = D\delta^d
\left(\vec{x}-\vec{x'}\right) \delta\left( t-t'\right)$, with $D$ constant, and
$d$ is the dimension of the substrate. In the following sections, we will study
those models in $2+1$ dimensions (in $1+1$ dimensions, some exact results and
highly accurate numerical results are already available).

The DT and the CRSOS models in $1+1$ dimensions are in the class of the
nonlinear fourth order growth equation \cite{predota,kotrla,huang,park}
\begin{equation}
{{\partial h}\over{\partial t}} =
\nu_4{\nabla}^4 h + \lambda_{22} {\nabla}^2 {\left( \nabla h\right) }^2 + \eta
(\vec{x},t) ,
\label{vlds}
\end{equation}
where $\nu_4$ and $\lambda_{22}$ are constants. This is also known
as Villain-Lai-Das Sarma (VLDS) equation \cite{villain,laidassarma}, and the
discrete models are said to belong to the VLDS class. The DT model is
particularly interesting in $1+1$ dimensions due to the presence of strong
finite-size corrections which leads to exponents estimates far from the VLDS
values (although noise reduction methods are able to provide estimates closer
to the VLDS ones \cite{punyindu}).

For all models, only steady state properties will be studied in this work. The
roughness distributions are typically obtained from ${10}^7$ to ${10}^8$
different configurations, and height distributions from a number of
configurations which is larger by a factor $L^2$ (the square of the lattice
size). 

\section{Scaling of roughness distributions}
\label{roughnessdistr}

In order to compare the roughness distributions of the above models, we use two
of them as representatives of their growth classes: the RSOS model with
${\Delta H}_{max}=1$, whose distribution hereafter will be referred as the
RSOS/KPZ one ($2+1$ dimensions, $L=256$), and the CRSOS model with ${\Delta
H}_{max}=1$, whose distribution hereafter will be referred as the CRSOS/VLDS
one ($1+1$ dimensions, $L=256$). The use of these
models data as standards is possible because previous work have already shown
that they have very small finite-size dependence \cite{distrib}.

The original BD model illustrates the need of choosing a suitable scaling
relation for the analysis of roughness distributions from small system sizes. In
Figs. 2a, 2b and 2c we
show its steady state distributions in $2+1$ dimensions obtained from $L=128$
to $L=512$ and scaled according to Eqs. (\ref{deff}), (\ref{defg}) and
(\ref{defh}), respectively. For comparison, the RSOS/KPZ scaled distributions
are also shown. In Fig. 2c, $C=12$ is used to match the peaks of the curves for
BD and for RSOS/KPZ.

It is clear that the scaling with the rms fluctuation $\sigma$ (Eq. \ref{defg}) 
is superior if compared with the scaling with Eq. (\ref{deff}). Indeed, there
is no data collapse in Fig. 2a, thus that plot is unable to confirm the KPZ
scaling of the BD model. From the discussion of Sec. 2, the deviations may be
related to finite-size corrections in the ratio $r$, and our estimates of $r$
for different sizes $L$ confirm this large size-dependence.

However, the quality of the data collapse using Eq. (\ref{defh}) is
similar to the one using Eq. (\ref{defg}). This result suggests that the
assumptions leading to the proposal of  Eq. (\ref{defh}) are reasonable, i. e.
the roughness of each configuration is a constant plus a KPZ-distributed
fluctuating part (Sec. II). It also gives an estimate of the intrinsic
width, which is the constant $C=12$ (Fig. 2c).

The scaled distributions of the grain deposition model with $l=16$ in lattice
sizes $L=2048$ and $L=4096$ are shown in Figs. 3a, 3b and 3c, considering Eqs.
(\ref{deff}), (\ref{defg}) and (\ref{defh}), respectively. The RSOS/KPZ
distribution is also shown. Deviations
from the scaling relation (\ref{deff}) are also large here, but using Eqs. 
(\ref{defg}) and (\ref{defh}) we obtain good data collapse with those large
lattice sizes. Similar results were obtained with an extended model where the
grains could have Poisson-distributed sizes, with an average size
$\overline{l}=16$ (the aggregation mechanism was the same of Fig. 1).

In BD and in this grain deposition model we observe
deviations from data collapse with sizes smaller than those shown in Figs.
3a-c. The deviations for a certain lattice size $L$ become smaller when the
grain size decreases, thus good data collapse is obtained for intermediate
values of $l$ ($l=2$ to $l=8$) with lattice sizes $L$ intermediate between
those in Figs. 2a-c and 3a-c. Typically, significant deviations appear only for
$L\leq 64l$ ($l=1$ for BD), probably due to further corrections to scaling,
whose effects are reduced as $L$ increases.

Our results for this group of ballistic-like models suggest that the
roughness is actually distributed in a KPZ-like form except for an additional
constant, which may be interpreted as the intrinsic width. Additional support to
the constant correction term in the scaling of $\langle w_2\rangle$ (Eq.
\ref{fvsat}) is given in Fig. 4, where we plot $\left( \langle w_2\rangle -
C\right) /L^{2\alpha}$ versus $L$, using $\alpha =0.39$ \cite{marinari,kpz2d}
and the values of $C$ obtained from collapse of roughness distributions. The
saturation value of $\left( \langle w_2\rangle - C\right) /L^{2\alpha}$ for 
each model gives an
estimate of the amplitude $A$ in Eq. (\ref{fvsat}). These estimates and the
correction constants $C$ are shown in Table I, where we also present the
corresponding values of both terms on the right-hand side of Eq. (\ref{fvsat})
for the largest lattice size used for each model ($L_{MAX}$). The ratios between 
the leading correction ($C$) and the dominant term $AL_{MAX}^{2\alpha}$ are 
very large for all model, in some cases exceeding $50\%$, which quantitatively 
confirm the strong finite-size effects.

However, it is important to notice that the scaling with Eq. (\ref{defh})
advances over Eq. (\ref{defg}) because
it not only confirms the universality class of the model (by data collapse) but
also provides evidence on the leading finite-size correction to the roughness
scaling. These results also advance over previous work where the intrinsic
width was identified only as a correction to scaling of the average roughness
\cite{stauffer,kertesz,tammaro,chavez,moro}.

The intrinsic width is usually associated to steep surface features, such as
large local slopes, holes and overhangs. In order to test this hypothesis, it
is interesting to consider other models which also have large local slopes.
This is the case, for instance, of the RSOS models with large ${\Delta
H}_{max}$.

In Figs. 5a and 5b we show the roughness distributions for the RSOS models
with ${\Delta H}_{max}=10$ and ${\Delta H}_{max}=20$ scaled according to Eqs.
(\ref{deff}) and (\ref{defg}), respectively. Both plots show excellent data
collapse with very small lattice sizes, such as $L=32$. Thus, there is no
correction term similar to the intrinsic width or, equivalently, $C\approx 0$
in Eq. (\ref{defh}) for the RSOS models.

Finally, we analyze the DT model in $1+1$ dimensions. In Figs. 6a
and 6b we show the roughness distributions for that model in lattice sizes
$L=64$ and $L=128$, scaled according to Eqs. (\ref{deff}) and (\ref{defg}).
No significant finite-size effect is found in the scaling with the fluctuation
$\sigma$ (Fig. 6b). However, both plots show that the distribution of the DT
model is very different from the CRSOS/VLDS distribution in $1+1$ dimensions
($L=256$) \cite{distrib}. Moreover, there is no evidence that
the former will converge to the CRSOS/VLDS curve as $L$ increases.

We observe the same large deviations when the DT distributions are scaled
according to  Eq.(\ref{defh}) for several values of the constant $C$. Thus, the
assumption of an intrinsic width is not sufficient to represent the main
finite-size effects in that model, at least in the accessible range for
simulation, which is $L\sim {10}^2$.

However, previous work showed that simulation of noise-reduced DT models
provides estimates of scaling exponents closer to the VLDS values
\cite{punyindu}. Thus, we also calculated the DT distributions using
noise-reduction parameters $m=10$, $m=20$ and $m=30$. Here, $m$ is the number of
times that a certain column has to be chosen for aggregation of a new particle
before the aggregation actually takes place \cite{punyindu}. In Fig. 6c we show
the scaled
distributions for these models in $L=64$, which converge to the CRSOS/VLDS curve
as $m$ increases (although reasonable data collapse is not obtained in the
tails).

This noise-reduction scheme is not expected to change the universality class of
the process because symmetries are preserved (see e. g. the discussion in Ref.
\protect\cite{hagston}). Thus, the above changes in the distributions when the
parameter $m$ is increased (Fig. 6c) are evidence that the asymptotic behavior
of the DT distribution was not attained yet. Consequently, the above
discrepancies with the CRSOS/VLDS curve do not mean that the DT model does not
belong to the VLDS class, but only that the finite-size effects on the
roughness distributions have much more complex forms than those suggested here.
Indeed, we recall that Ref. \protect\cite{predota} derived the VLDS equation
from the master equation of the DT model in $1+1$ dimension. 

In the following Sections, we will show that finite-size
corrections obtained here in the roughness distributions are accompannied by
corrections in other quantities.

\section{The role of the local slopes}

Here we characterize the interfaces at small lengthscales by the average
squared height difference between nearest neighbor columns, $\langle {{\delta
h}_{nn}}^2 \rangle$, also calculated at the steady states. Our main aim is to
test the hypothesis that the intrinsic width is related to local height
fluctuations of the interface.

In the ballistic-like models, $\langle {{\delta h}_{nn}}^2 \rangle$ has a
significant finite-size dependence for small lattice sizes. However, it is
approximately constant in the largest lattice sizes analyzed for each $l$ in
Sec. III (the same sizes used to collapse the roughness distributions).
However, in the RSOS models $\langle {{\delta h}_{nn}}^2 \rangle$ attains a
saturation value for very small system sizes.

In Fig. 7 we plot the correction term $C$ obtained in the scaling with Eq.
\ref{defh} (i. e. the intrinsic width) as a function of the large $L$ value of 
$\langle {{\delta h}_{nn}}^2 \rangle$ for both groups of models. The result for
the ballistic-like models could naively suggest that the intrinsic width was
only related to local height fluctuations. However, the result for the RSOS
models show this interpretation is not valid in general: $C\approx 0$ in that
case and the values of $\langle {{\delta h}_{nn}}^2 \rangle$ are in the same
range of those for ballistic-like models with $l\leq 4$.

In the DT model in $1+1$ dimensions, we observe a rapid increase of the average
local slope with the lattice size: $\langle {{\delta h}_{nn}}^2 \rangle \approx
22$ for $L=32$, $\langle {{\delta h}_{nn}}^2 \rangle \approx 58$ for $L=64$,
and $\langle {{\delta h}_{nn}}^2 \rangle \approx 139$ for $L=128$. A comparison
with $\langle {{\delta h}_{nn}}^2 \rangle$ for the other models shown in Fig. 7
is not possible because we are not able to perform a reliable extrapolation of
these data to $L\to\infty$. On the other hand, these data confirm that the DT
model is very far from its asymptotic limit (the continuum VLDS model) in this
range of lattice sizes.

The above results suggest that the presence of large fluctuations at small
lengthscales is insufficient to explain the intrinsic width as the main
correction term of roughness scaling in the ballistic-like models, since this
is also a property of models which have much weaker scaling corrections (the
RSOS ones). On the other hand, we observe that finite-size effects on roughness
distributions are accompannied by the same type of effect on $\langle {{\delta
h}_{nn}}^2 \rangle$.

\section{Finite-size effects in height distributions}

Finite-size effects in height distributions are typically larger than those in
roughness distributions, as illustrated for KPZ models in Ref.
\protect\cite{localKPZ}. Thus, in order to study finite-size effects on
steady state height distributions, the best procedure is to compare their
skewness $S$ and kurtosis $Q$ in different lattice sizes  (other previous works
which aimed at characterizing height distributions also focused on those
quantities - see e.g. Refs. \protect\cite{chin,shim,marinari,kpz2d}). They are
defined from the moments of the height distributions, $W_n \equiv {\left<
\overline{ {\left( h-\overline{h}\right) }^n } \right> }$, as
\begin{equation}
S \equiv {{W_3}\over{{W_2}^{3/2}}}
\label{defskew}
\end{equation}
and
\begin{equation}
Q \equiv {{W_4}\over{{W_2}^{2}}}-3 .
\label{defkurt}
\end{equation}

The sign of the skewness indicates which type of singular feature is dominant at
the interface: deep valleys or grooves lead to negative $S$, while sharp peaks
give positive $S$. In KPZ models the sign of $S$ is the same of the coefficient
of the nonlinear term $\lambda$ of the corresponding KPZ equation (see e. g.
the discussion in Ref. \protect\cite{kpz2d}). Thus, since ballistic-like models
correspond to positive $\lambda$ and RSOS models to negative $\lambda$, we must
compare values of $S$ from the former with $-S$ from the latter.

In Figs. 8a and 8b we show $\pm S$ and $Q$ versus $1/L^{1/2}$, respectively, for
BD, for the grain deposition model with $l=8$ and for the RSOS model with
${\Delta H}_{max}=10$ ($-S$ only for the RSOS model). Data for BD in the
largest sizes were extracted from Ref. \protect\cite{kpz2d}. In Figs. 8a and 8b,
the variable $1/L^{1/2}$ was used in the abscissa because it is the one which
provides the best linear fits among other variables of the form $1/L^\Delta$,
with one digit variation in $\Delta$. Since the asymptotic values of $S$ and
$Q$ are not known exactly, as well as their finite-size corrections, such
numerical extrapolations are necessary.

We observe very weak finite-size corrections in the RSOS data, so that a simple
extrapolation procedure to $L\to\infty$ ($1/L^{1/2}\to 0$) leads to an
asymptotic estimate of $S$ in agreement with the best known value $|S|=0.26\pm
0.01$ of the KPZ class \cite{marinari,kpz2d}.
However, there are large finite-size corrections in the
data for the BD model. It is remarkable that up to $L\approx 500$, the skewness
of BD is negative, while the values for larger $L$ are positive. The
finite-size dependence of the data for the grain deposition model is much
stronger: only for $L\approx 4096$ the skewness crosses over from negative to
positive values.

The change in sign of $S$ of the ballistic-like models shows that the typical
steady state configurations
for small $L$ are very different from the typical ones for large $L$. In other
words, the steady states for small $L$ are not representative of the asymptotic
limit of the model (large $L$ and $t$), which is the KPZ limit. For the grain
deposition model with $l=16$, even the steady state for $L=4096$ has $S\approx
0$ and, consequently, is very far from the KPZ limit.

Despite these problems, the data for the BD ballistic-like models indicate that
$S$ is positive when $L\to\infty$. Asymptotic estimates (not very accurate) of
$S$ can be obtained from extrapolations of those data; they are not very
distant from the best known values for KPZ in $2+1$ dimensions. However, the
finite-size effects on roughness distributions are certainly smaller than those
found in height distributions. Indeed, the good data collapse in Figs. 2b, 2c,
3b and 3c is reflected in values of skewness (of the roughness distributions)
close to the RSOS/KPZ curve.

The analysis of the kurtosis of the above models (Fig. 8b) leads to the same
conclusions, although there the finite-size effects are much larger, as well as
the error bars of the data.

The values of $S$ and $Q$ for the DT and the CRSOS models are shown in Figs. 9a
and 9b as a function of $1/L^{1/2}$. They were estimated up to $L=256$ for the
DT model (this work) and up to $L=2048$ for the CRSOS model (Ref.
\protect\cite{crsos2}), with a relatively low accuracy for the largest $L$
data. Here the variable in the abscissa was chosen by analogy with Fig. 8, but
not for extrapolation purposes.

The asymptotic skewness of the VLDS class,
$|S|=0.32\pm 0.02$, can be obtained from extrapolation of the CRSOS data to
$L\to\infty$ \cite{crsos2}. However, it is clear that the data for the DT
model in Fig. 9a do not converge to the same value, even if the sign of $S$ is
changed (the
actual sign of $S$ is related to that of $\lambda_{22}$ in Eq. \ref{vlds}).
There is also no evidence that the kurtosis of the DT model will converge to
the same value of the CRSOS model. Thus, the height distributions of
the DT model up to $L=256$ are not representative of the VLDS class, even if we
account for simple finite-size corrections.

The overall conclusion from the above results is that small or large finite-size
corrections are simultaneously present in roughness and height distributions,
but the convergence of roughness distributions to the asymptotic limit is
certainly much better when it occurs, as illustrated with the ballistic-like
models. The amplitude ratios of height distributions are interesting to show
that, when the corrections are large, the finite-size configurations of hills
and valleys are not representative of the universality class of the model. We
are not able to explain why the roughness distributions (for large $L$)
represent the correct class of the model when the height distributions do not,
but certainly there is no contradiction in this finding because we are
comparing distributions of local and global quantities, whose scaling
properties may be very different.

\section{Conclusion}
\label{conclusion}

For various interface growth models, height and roughness distributions and
average square neighboring heights differences $\langle {{\delta h}_{nn}}^2
\rangle$ were numerically calculated in the steady states, in order to analyze
their finite-size effects.

First, we considered a group of ballistic-like deposition models in which the
most typical scaling relation for roughness distributions (Eq. \ref{deff})
fails to provide collapse of finite-size systems data. On the other hand, good
data collapse was obtained with a scaling relation involving the roughness
fluctuation (Eq. \ref{defg}) and with an alternative relation which includes the
effect of the intrinsic width. This shows that this constant term is the main
correction to the asymptotic KPZ distribution, thus extending previous work
which suggested that quantity only as a correction to the scaling of the
average roughness. The comparison with other models with large $\langle
{{\delta h}_{nn}}^2 \rangle$ (the RSOS ones) show that the average local slope
is not sufficient to explain the intrinsic width, in contrast to previous
interpretation. Height distributions for those models also show strong
finite-size effects, although results for sufficiently large lattice sizes
indicate convergence of amplitude ratios to the expected asymptotic
values.

We also studied the model of Das Sarma and Tamborenea (DT) in $1+1$
dimensions. Its roughness distributions up to size $L=128$ are very different
from the ones representative of the asymptotic VLDS class. Strong finite-size
effects are also observed in skewness and kurtosis of height distributions and
in $\langle {{\delta h}_{nn}}^2 \rangle$, with no evidence of convergence to
reliable asymptotic values.

Our results show that the deviations from finite-size scaling in the roughness
distributions are usually accompanied by significant finite-size dependence of
dimensionless amplitude ratios of moments of height distributions and of
average local slopes. One important point is that this conclusion is not
restricted to the KPZ class. The analysis of simple quantities, such as $\langle
{{\delta h}_{nn}}^2 \rangle$, is useful to search for finite-size effects and
may reveal slow convergence to a continuum limit. In some situations (e. g. the
DT model here), this convergence may be so slow that no one of the quantities
analyzed here are able to show the correct class of the growth process. In
the other cases, roughness distributions are certainly better than other
quantities (including scaling exponents) to classify the process.

\acknowledgments

TJO acknowledges support from CNPq and FDAAR acknowledges support from CNPq and
FAPERJ (Brazilian agencies).


\vskip 5cm

\begin{table}[!h]
\caption{For each ballistic-like model, the amplitude $A$ obtained from the
saturation of $\left( \langle w_2\rangle -C\right) /L^{2\alpha}$, the constant
$C$ used to scale roughness distributions, the maximal dominant term of average
roughness scaling ($AL_{MAX}^{2\alpha}$) and the relative importance of the
correction term.}
\vskip 0.2cm
\halign to \hsize
{\hfil#\hfil&\hfil#\hfil&\hfil#\hfil&\hfil#\hfil&\hfil#\hfil&\hfil#\hfil\cr
Model & \ $A$ & \ $C$ & \ $AL_{MAX}^{2\alpha}$ & \
$\frac{C}{AL_{MAX}^{2\alpha}}$ \cr
BD & \ $0.35$ & \ $12$ & \ $45$ & \ $27\%$ \cr
$l=2$ & \ $3.4$ & \ $120$ & \ $440$ & \ $27\%$ \cr
$l=4$ & \ $6.6$ & \ $630$ & \ $1470$ & \ $43\%$ \cr
$l=8$ & \ $14.9$ & \ $2950$ & \ $5700$ & \ $52\%$ \cr 
$l=16$ & \ $34.0$ & \ $12800$ & \ $22340$ & \ $57\%$ \cr  }
\label{table1}
\end{table}

\newpage

\begin{figure}[!h]
\includegraphics[width=13cm]{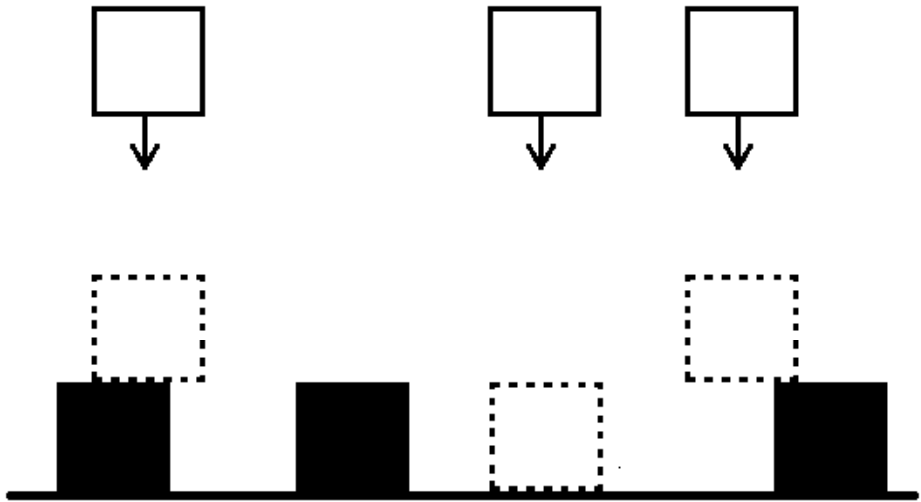}
\caption{Illustration of the deposition rules of the grain aggregation
model. Shaded squares are previously deposited grains, open squares are
incident grains and open dashed squares show their final aggregation positions.}
\label{fig1}
\end{figure}

\begin{figure}[!h]
\includegraphics[width=10cm]{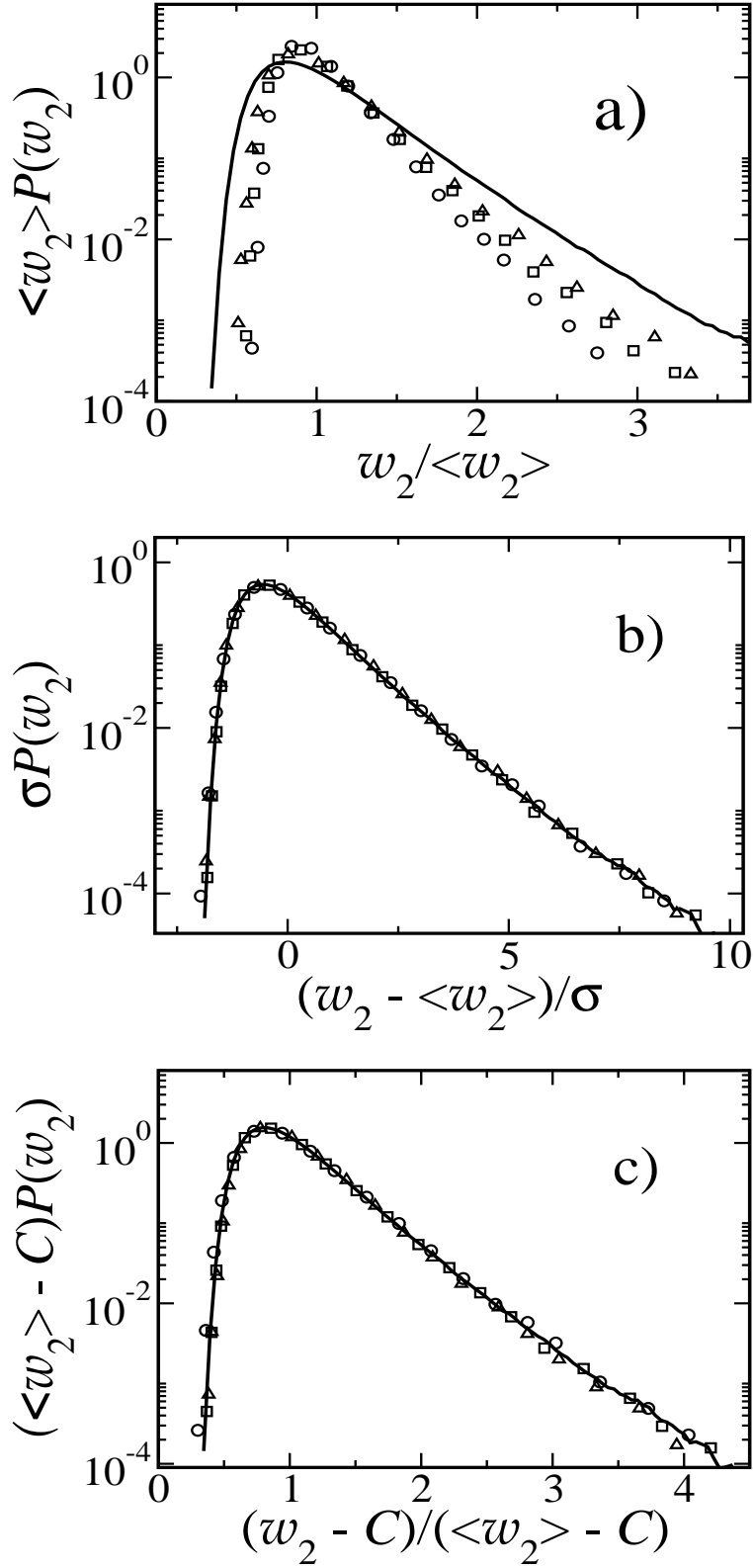}
\caption{Steady state normalized square roughness distributions of the BD model
for lattice sizes $L=128$ (circles), $L=256$ (squares) and $L=512$ (triangles),
scaled according to: (a) Eq. (\ref{deff}), (b) Eq. (\ref{defg}) and (c) Eq.
(\ref{defh}). In each plot, the solid curve is the RSOS/KPZ distribution scaled
accordingly.}
\label{fig2}
\end{figure}

\begin{figure}[!h]
\includegraphics[width=10cm]{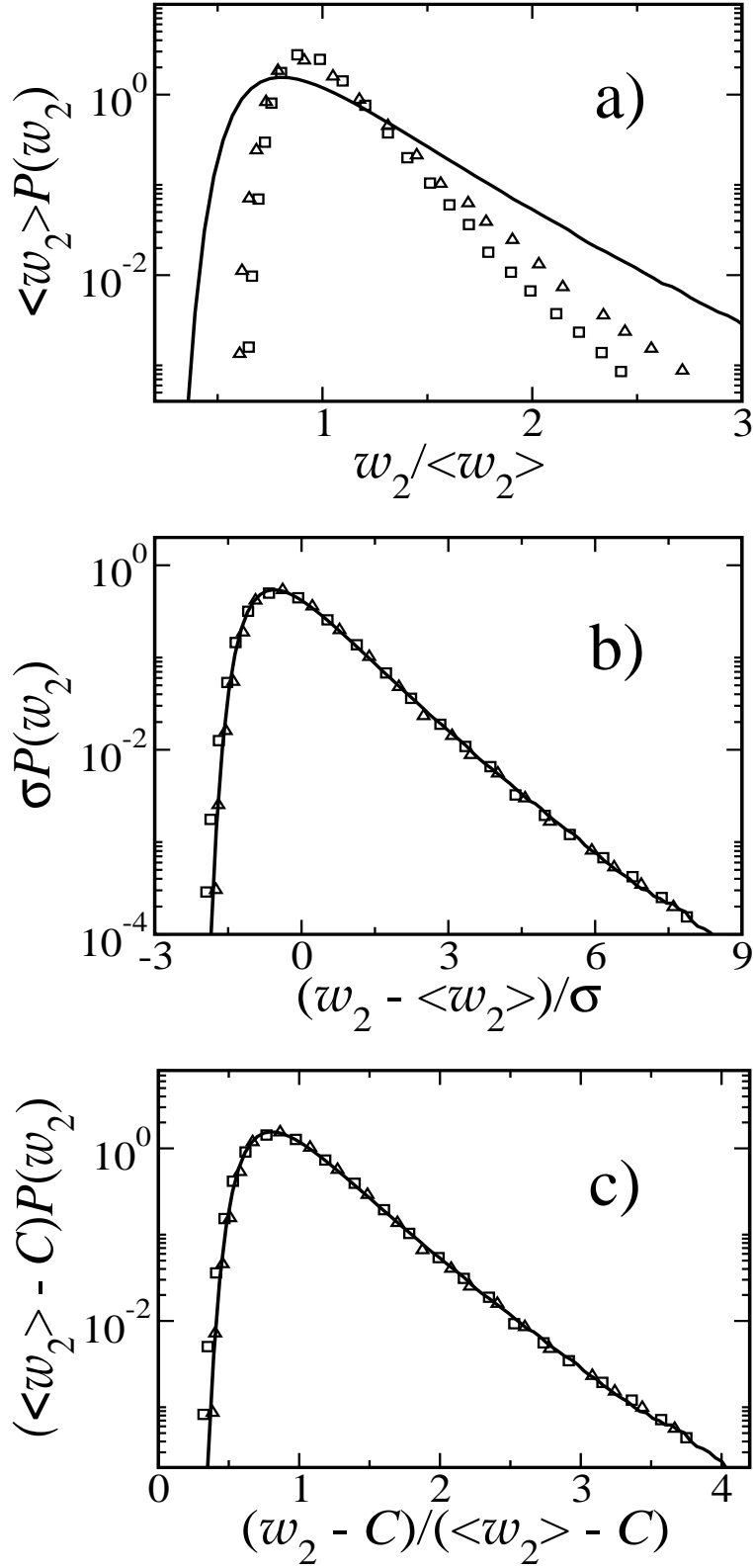}
\caption{Steady state square roughness distributions of the grain
deposition model with $l=16$ and lattice sizes $L=2048$ (squares) and $L=4096$
(triangles), scaled according to: (a) Eq. (\ref{deff}), (b) Eq. (\ref{defg}) and
(c) Eq. (\ref{defh}). In each plot, the solid curve is the RSOS/KPZ distribution
scaled accordingly.}
\label{fig3}
\end{figure}

\begin{figure}[!h]
\includegraphics[width=10cm]{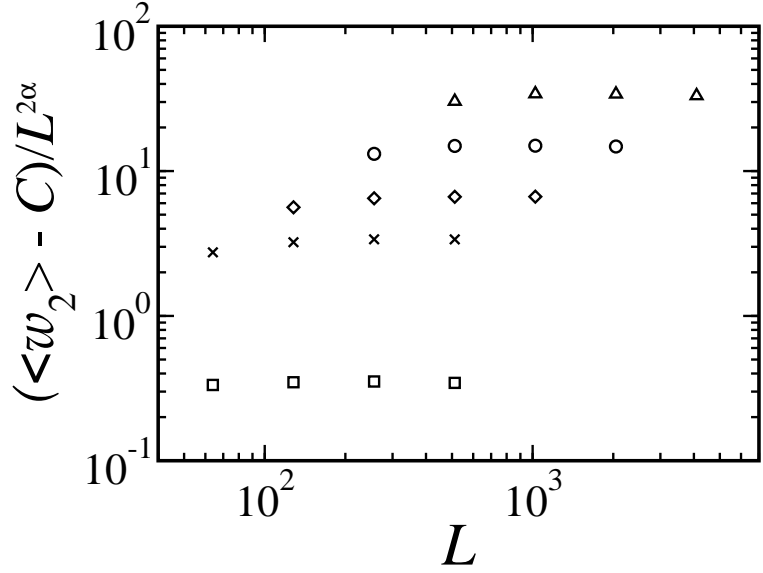}
\caption{$\left( \langle w_2\rangle -C\right) /L^{2\alpha}$ versus lattice size
$L$ for the ballistic like models, where $C$ is the constant used to fit
roughness distributions to the new scaling relation. Symbols correspond to the
ballistic deposition model (squares) and the grain deposition models with $l=2$
(crosses), $l=4$ (diamonds), $l=8$ (circles) and $l=16$ (triangles).}
\label{fig4}
\end{figure}

\begin{figure}[!h]
\includegraphics[width=10cm]{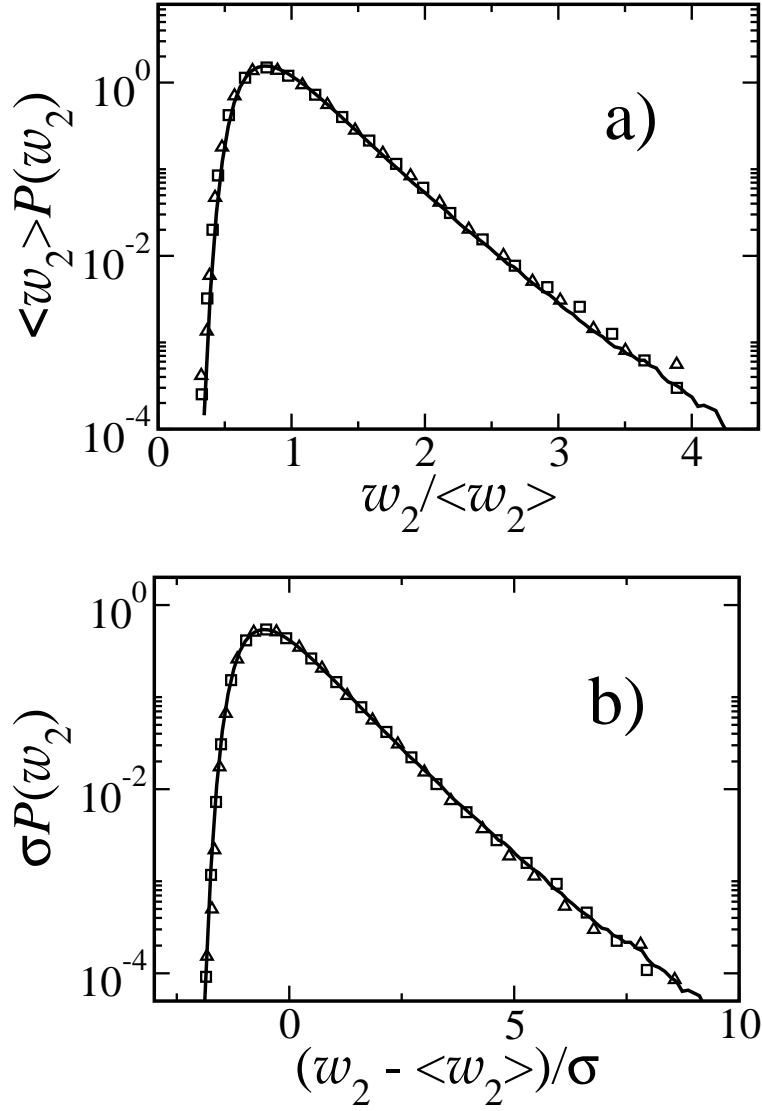}
\caption{Steady state normalized square roughness distributions of the
RSOS model with $\Delta H_{max}=10$ (squares) and $\Delta
H_{max}=20$ (triangles) scaled according to: (a) Eq. (\ref{deff}) and (b) Eq.
(\ref{defg}). In each plot, the solid curve is the RSOS/KPZ distribution scaled
accordingly.}
\label{fig5}
\end{figure}

\begin{figure}[!h]
\includegraphics[width=10cm]{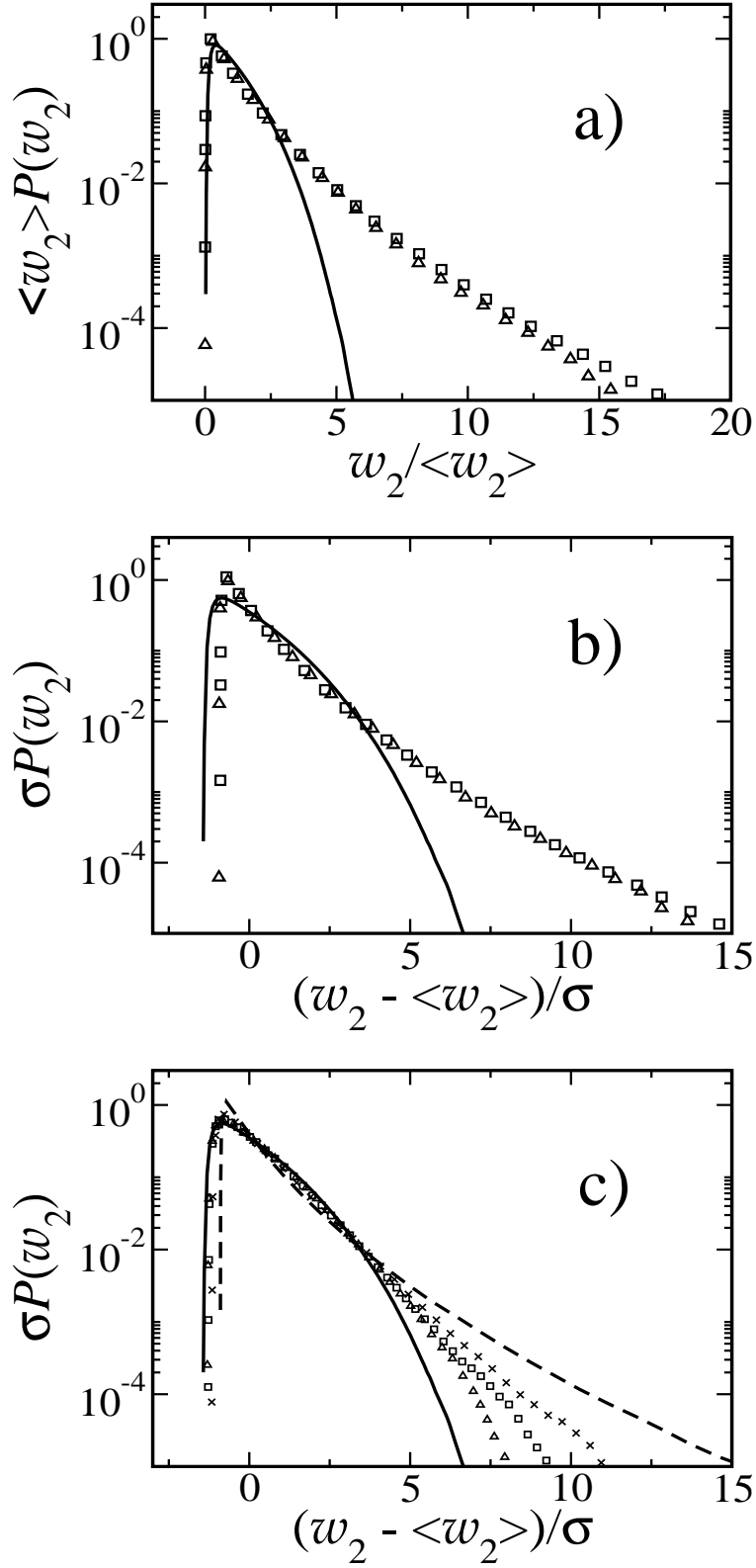}
\caption{(a), (b): steady state normalized square roughness distributions of the
DT model in lattice sizes $L=64$ (squares) and $L=128$ (triangles) scaled
according to Eq. (\ref{deff}) and Eq. (\ref{defg}), respectively; (c)
distributions for the original and the noise-reduced DT models in lattice size
$L=64$: $m=1$ (original model - dashed line), $m=10$ (crosses), $m=20$
(squares) and $m=30$ (triangles). In all plots, the solid curve is
the CRSOS/VLDS distribution scaled accordingly.}
\label{fig6}
\end{figure}

\begin{figure}[!h]
\includegraphics[width=10cm]{Fig7.eps}
\caption{The average square nearest neighbor height difference $\langle 
{{\delta h}_{nn}}^2 \rangle$ of the ballistic-like models (open circles) and of
the RSOS models (triangles).}
\label{fig7}
\end{figure}

\begin{figure}[!h]
\includegraphics[width=10cm]{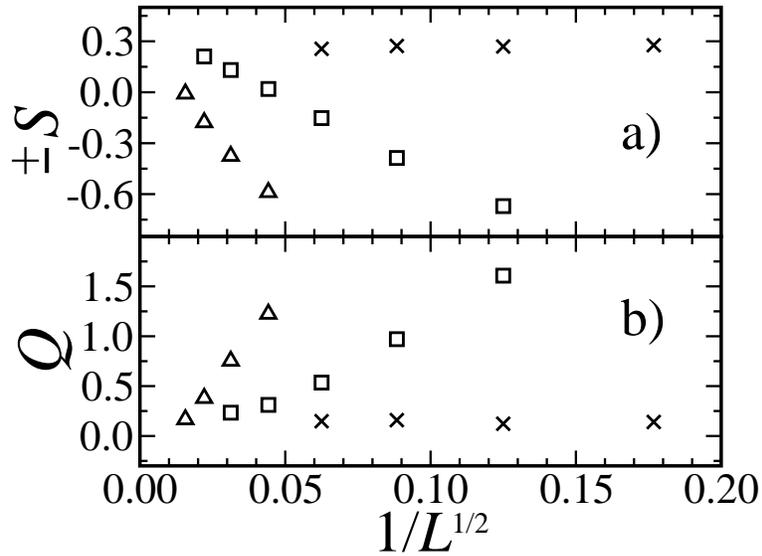}
\caption{Skewness $\pm S$ (a) and kurtosis $Q$ (b) of height distributions for
the RSOS model with $\Delta H_{max}=10$ (crosses), the BD model
(squares) and the grain deposition model with $l=8$ (triangles). $-S$ is
plotted only for the RSOS model.}
\label{fig8}
\end{figure}

\begin{figure}[!h]
\includegraphics[width=10cm]{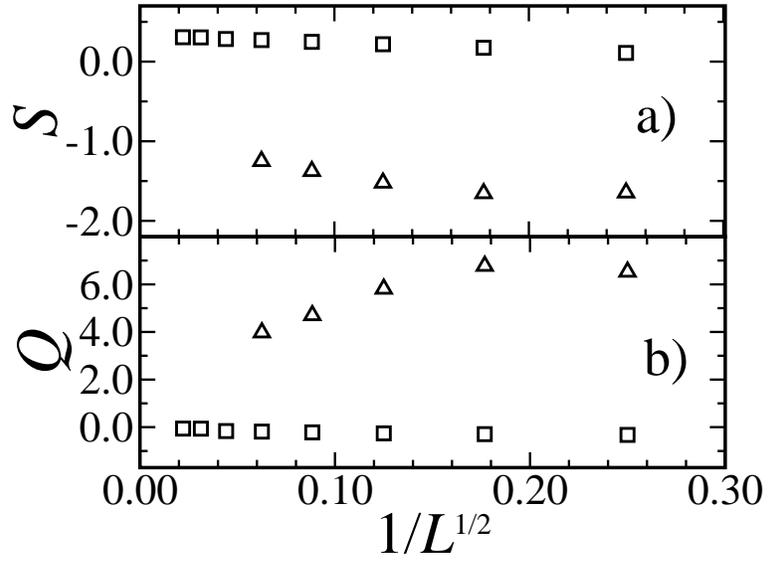}
\caption{Skewness $S$ (a) and kurtosis $Q$ (b) of height distributions for
CRSOS model (squares) and DT model (triangles).}
\label{fig9}
\end{figure}

\end{document}